\documentclass[twocolumn,prb,aps,superscriptaddress,altaffilletter,showpacs]{revtex4-1}

\usepackage{amsmath,amssymb}
\usepackage{graphicx}
\usepackage{xcolor}

\begin{document}


\title{Thermoelectric power quantum oscillations in the ferromagnet UGe$_{2}$}

\author{A. Palacio Morales}
\altaffiliation{Present address: Institute of Applied Physics and and Interdisciplinary Nanoscience Center Hamburg, University of Hamburg, 20355 Hamburg, Germany}
\affiliation{Univ. Grenoble Alpes, INAC-SPSMS, F-38000 Grenoble, France}
\affiliation{CEA, INAC-SPSMS, F-38000 Grenoble, France}
\author{A. Pourret}
\email{alexandre.pourret@cea.fr}
\affiliation{Univ. Grenoble Alpes, INAC-SPSMS, F-38000 Grenoble, France}
\affiliation{CEA, INAC-SPSMS, F-38000 Grenoble, France}
\author{G. Knebel}
\email{georg.knebel@cea.fr}
\affiliation{Univ. Grenoble Alpes, INAC-SPSMS, F-38000 Grenoble, France}
\affiliation{CEA, INAC-SPSMS, F-38000 Grenoble, France}
\author{G. Bastien}
\affiliation{Univ. Grenoble Alpes, INAC-SPSMS, F-38000 Grenoble, France}
\affiliation{CEA, INAC-SPSMS, F-38000 Grenoble, France}
\author{V. Taufour}
\altaffiliation{Present permanent address: Ames Laboratory, US Department of Energy, Iowa State University, Ames, Iowa 50011, USA}
\affiliation{Univ. Grenoble Alpes, INAC-SPSMS, F-38000 Grenoble, France}
\affiliation{CEA, INAC-SPSMS, F-38000 Grenoble, France}
\author{D. Aoki}
\affiliation{Univ. Grenoble Alpes, INAC-SPSMS, F-38000 Grenoble, France}
\affiliation{CEA, INAC-SPSMS, F-38000 Grenoble, France}
\affiliation{IMR, Tohoku University, Oarai, Ibaraki 311-1313, Japan}
\author{H. Yamagami}
\affiliation{Condensed Matter Science Division, Japan Atomic Energy Agency, Sayo, Hyogo 679-5148, Japan}
\affiliation{Department of Physics, Faculty of Science, Kyoto Sangyo University, Kyoto 603-8555, Japan}
\author{J. Flouquet}
\affiliation{Univ. Grenoble Alpes, INAC-SPSMS, F-38000 Grenoble, France}
\affiliation{CEA, INAC-SPSMS, F-38000 Grenoble, France}

\date{\today}

\begin{abstract}
We present thermoelectric power and resistivity measurements in the ferromagnet UGe$_2$ as a function of temperature and magnetic field. At low temperature, huge quantum oscillations are observed in the thermoelectric power as a function of the magnetic field applied along the $a$ axis.
The frequencies of the extreme orbits are determined and an analysis of the cyclotron masses is performed following different theoretical approaches for quantum oscillations detected in the thermoelectric power. They are compared to those obtained by Shubnikov-de Haas experiments on the same crystal and  previous de Haas-van Alphen experiments.  The agreement of the different probes confirms thermoelectric power as an excellent  probe to extract simultaneously both microscopic and macroscopic information on the Fermi-surface properties. Band-structure calculations of UGe$_2$ in the ferromagnetic state are compared to the experiment. 
\end{abstract}

\pacs{71.18.+y, 72.15.Jf, 71.27.+a, 74.70.Tx}

\maketitle

\section{Introduction}\label{Intro}

The determination of the Fermi surface is a fundamental key ingredient for the understanding of the physical properties of metals. Quantum oscillation experiments, such as de Haas van Alphen (dHvA) or Shubnikov de Haas (SdH), are powerful experimental tools to probe the topology of the Fermi surface by detecting the extreme orbits of the electrons and determining their effective masses. These effects can be observed in high quality samples at low temperatures and in high magnetic fields. Both, dHvA and SdH, are standard methods to determine the Fermi surface properties in strongly correlated electron systems. 
While in simple metals like Al, Be, In, or Zn  quantum oscillations in the thermoelectric power (TEP) have been reported already more than forty years ago (see e.~g.~Ref.\onlinecite{Papastaikoudis1979} and references therein), their observation in strongly correlated electron systems is rather new \cite{Laliberte2011, Hodovanets2013, Boukahil2014}. 

The theoretical understanding of  TEP quantum oscillations is different from that of the more conventional thermodynamic or transport measurements, dHvA and SdH. In thermodynamic probes like specific heat, magnetization, but also in SdH, the oscillations are directly linked to the oscillations of the density of states  $N(\epsilon)$ and the amplitude of the oscillations follows the Lifshitz-Kosevich theory \cite{Shoenberg1984}. The unique aspect of TEP is that it contains information on both transport and thermodynamic properties of the system which makes the interpretation rather difficult. In a metal the TEP depends on the logarithmic energy derivative of the electronic conductivity according to the Mott formula\cite{Mott1958}. It is thus dependent on the energy derivative of the density of states which is related to the electronic entropy  per charge carrier, but also on the energy dependence of the scattering time which is a transport property. Both contributions can contribute to the oscillatory part of the thermoelectric power under magnetic field.\cite{Ikeda2001}

In the present article we discuss the thermoelectric power in the ferromagnet UGe$_2$ with the main focus on the analysis of the observed quantum oscillations. UGe$_2$ has gained special attention as pressure induced superconductivity coexists with the ferromagnetic order below the critical pressure $p_c = 1.5$ GPa where the ferromagnetism is suppressed \cite{Saxena2000}. Compared to the other uranium based ferromagnetic superconductors, excellent single crystals of this system can be grown and thus the system plays a key role in the research of unconventional superconductivity. Furthermore, UGe$_2$ is now a paradigm for the study of ferromagnetic quantum criticality with the emergence of ferromagnetic wings structure \cite{Taufour2010, Kabeya2010, Kotegawa2011, Wysokinski2015}.

UGe$_2$ crystallizes in the orthorhombic ZrGa$_2$ type crystal structure (space group $Cmmm$) with antiphase zigzag chains of U atoms along the $a$ axis \cite{Oikawa1996, Boulet1997}. The crystal structure presents a strong magnetic anisotropy \cite{Menovsky1983, Onuki1992, Boulet1997, Sakon2007} with the magnetic moments aligned along the zigzag chains ($a$ axis) which is the easy magnetization axis.  At ambient pressure UGe$_2$ orders ferromagnetically at $T_C=52$~K. In the temperature range from 20~K to 35~K a cross-over from a high temperature weakly polarized FM1 to a low temperature strongly polarized ferromagnetic phase FM2 is observed in several quantities, but at ambient pressure no signature of a phase transition is observed \cite{Hardy2009}. 
As function of pressure the magnetic moment determined by bulk magnetization measurements jumps from $\mu_0=1.4 \mu_{B}/$U in the FM2 phase to $\mu_0=0.9 \mu_{B}/$U in the weakly polarized FM1 state by a first order transition at $p_x (0) \approx 1.2$~GPa at $T=0$.\cite{Pfleiderer2002} 
At finite temperature this first order transition ends at a critical end point ($T_{CEP} = 7$~K and $p_{CEP} = 1.16$~GPa) \cite{Taufour2010, Taufour2011}. Thus the crossover FM1 -- FM2 at zero pressure is reminiscent of the critical end point.
From measurements of the magnetic form factor by neutron scattering\cite{Kernavanois2001} and also from muon spin rotation and relaxation ($\mu$SR) \cite{Yaouanc2002, Sakarya2010} it has been concluded that the magnetic moment responsible for bulk magnetic properties is localized on the U site and the diffuse component arising from conduction electrons is small. Furthermore, a positron annihilation study suggests that the 5$f$-electron itinerant description does not apply to the paramagnetic phase of UGe$_2$.\cite{Biasini2003} In difference to this localized picture the observation of large cyclotron effective masses in quantum oscillation experiments \cite{Onuki1991,Satoh1992} shows that the 5$f$ electrons cannot be considered as fully localized, and  the nature of the U 5$f$ state is still controversial and might be described by a duality model with two-subset electronic systems of localized and itinerant character, respectively. \cite{Zwicknagl2003, Runge2004}  A dual nature of the 5$f$ electrons has been supported in Refs.~\onlinecite{Yaresko2005, Samsel-Czekala2011,  Troc2012}. Conversely, a recent angle-resolved photoelectron spectroscopy using soft X-rays suggests that the U 5$f$ electrons participate to quasi-particle bands and show an itinerant character in the paramagnetic (PM) state, even up to 120~K \cite{Fujimori2015}. The pressure induced superconductivity is discussed in an itinerant electron picture.\cite{Sandeman2003} Depending on the experimental probe localized or itinerant character will be revealed.

Previously, the Fermi surface of UGe$_2$ has been studied in detail by dHvA experiments. The main Fermi surfaces with heavy electron masses are highly corrugated but cylindrical along the $b$ axis \cite{Onuki1991, Satoh1992}. Large effective masses up to 25~$m_0$ have been observed for a field along the $b$ axis. Only small Fermi surface branches have been detected for a magnetic field along the $a$ axis. In high pressure experiments at each transition, from strongly polarized FM2 state to weakly polarized FM1, and from FM1 to PM state, abrupt Fermi surface changes have been observed \cite{Terashima2001,Haga2002, Settai2002, Terashima2002} with strong feedback on the magnetic and superconducting properties \cite{Sandeman2003}.

In this article we present quantum oscillations which have been observed for magnetic field $H \parallel a$ axis in the thermolelectric power. These will be compared to SdH experiments performed on the same single crystal. While huge quantum oscillations appear in the TEP, only very tiny oscillations could be observed in the magnetoresistance. Different approaches to analyze the strong TEP quantum oscillations will be discussed. 

Finally, we present new band structure calculations for a polarization along the $a$ axis and compare the observed quantum oscillation frequencies to the calculated Fermi surface. However, the agreement is still not satisfying indicating the difficulty of electronic structure calculations with 5$f$ bands contributions at the Fermi level notably for low symmetry crystals such as the orthorhombic UGe$_2$. In the class of discovered FM superconductors (UGe$_2$, URhGe, UIr, UCoGe) UGe$_2$ is the only easy case to grown high quality single crystals with residual resistivity ratio (RRR) above 100. So it is the only material where large parts of the Fermi surface can be determined by quantum oscillations and thus it is an excellent system to test reliability achieved to day in band structure calculation.

\section{Experimental details} \label{experiment}

Single crystals of UGe$_2$ are grown by the Czochralski method in a tetra-arc furnace, oriented by X-ray Laue diffraction and cut with a spark-cutter into bar-shaped samples. UGe$_2$ single crystals solidify out of a congruent melt and very high quality single crystals can be obtained. TEP measurements were performed applying a magnetic field along the $a$ axis and  thermal gradient $\nabla T$ along the $b$-axis. The demagnetization factor corresponding to the shape of the sample measured was evaluated to $D \approx 1$. As a consequence, the total magnetic field applied to the sample corresponds to the external magnetic field. The magnetoresistivity and TEP of two different crystals (named sample \#1 and \#2) have been measured. The residual resistivity ratio (RRR) of the studied crystals are around $300$.  From the observed quantum oscillations (see below) we can conclude that the quality of sample \#2 is higher than that of sample \#1.

The thermoelectric power measurements were performed at low temperatures down to $180$~mK and under magnetic fields up to $16$~T using a ``one heater and two thermometers'' setup. The thermometers have been calibrated under magnetic field up to 16~T down to 100~mK against a Germanium thermometer which is installed in the field compensated region of the superconducting magnet. Thermometers and heater are thermally decoupled from the sample holder by highly resistive manganin wires (200 $\Omega$/m). The temperature and field dependence of the TEP have been measured by averaging the TEP voltage during several minutes with and without thermal gradient. To observe quantum oscillations in the TEP, the field has been swept continuously upwards. A constant power was applied to the heater in order to obtain a thermal gradient during the field sweep. The thermoelectric voltage without thermal gradient was taken at the beginning and at the end of the sweep. 

Accurate resistivity measurements have been performed down to 30~mK and fields up to 13~T by a four point lock-in technique using a low temperature transformer to improve the signal to noise ratio. The maximal applied current was $I=100 \mu$A. For $H \parallel a$ the same single crystals than for the thermoelectric power have been used. Furthermore, magnetoresistivity measurements have been performed for $H \parallel b$ on different crystals. All data shown are obtained by sweeping the magnetic field upwards. 

\section{Results and discussion}

\subsection{Temperature Dependence of Thermoelectric Power}
\begin{figure}[t!]
\includegraphics[width=0.9\linewidth]{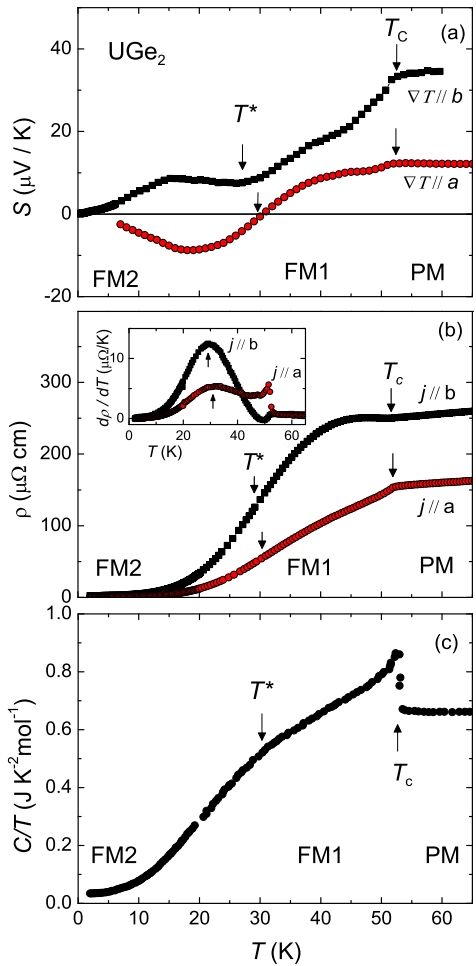}
\caption{\label{SvsSdHvsCvsT} (Color online). (a) TEP ($S$), (b) resistivity $\rho$ and (c) specific heat divided by temperature $C/T$ as a function of temperature at zero magnetic field. In TEP and resistivity measurements, the thermal gradient and current are applied along the $a$ and $b$ axis, respectively. The inset in (b) shows the derivative of the resistivity for both current directions. The arrows  indicate the position of $T_c = 52$~K and $T^\star\approx 30$~K in the different experimental probes.}
\end{figure}

Figure~\ref{SvsSdHvsCvsT} shows the temperature dependence of (a) the thermoelectric power $S$, (b) resistivity $\rho$ and (c) specific heat divided by temperature, $C/T$, of UGe$_2$ at zero magnetic field. At the Curie temperature $T_C$ all probes indicate distinct anomalies, while the anomalies at $T^\star$ are less pronounced. 

The temperature dependence of $S$ for heat current $\nabla T \parallel b$ is very similar to that reported in Ref.~\onlinecite{Onuki1992} obtained on a polycrystalline sample. At $T_C$ a sharp kink marks the ferromagnetic transition. On further cooling a small bump at $T\sim 40$ K and a well defined anomaly at $T^\star$ (corresponding to a minimum) and a small bump at $T\sim 15$~K can be observed. In difference, for a heat current $\nabla T \parallel a$ the thermoelectric power at $T_C$ is more than twice smaller and remarkably, $S(T)$ changes sign at $T^\star$ and $\mid S\mid$ has a maximum at 20~K.

\begin{figure}[t!]
\includegraphics[width=0.9\linewidth]{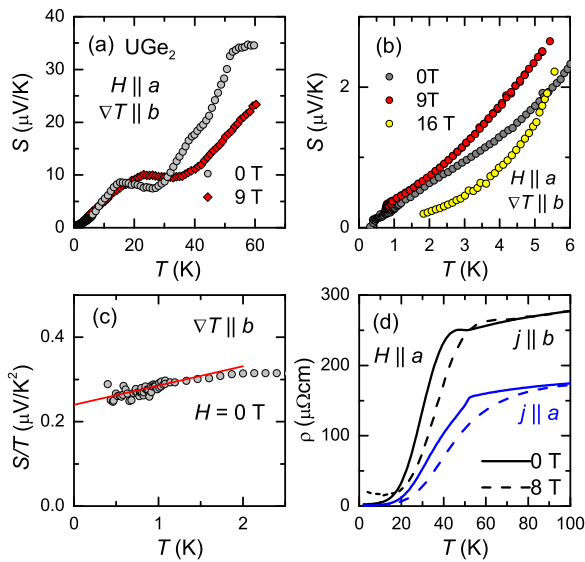}
\caption{\label{fig2} (Color online). (a) Temperature dependence of $S(T)$ for $H=0$ and $H=9$~T for the transverse configuration. (b) Zoom  on $S(T)$  below 6~K for thermal gradient $\nabla T \parallel b$ axis at different magnetic field. (c) $S/T$ at zero field in the low temperature limit. (d)
Temperature dependence of the resistivity at $H=0$ (solid lines) and 8~T (dashed lines) along the $a$ axis for transverse ($j \parallel b$) and longitudinal ($j \parallel a$) configuration. }
\end{figure}

In the resistivity  (Fig.~\ref{SvsSdHvsCvsT}(b)) a small hump of $\rho $ at $T_C$ for $j \parallel b$ appears just below $T_C$ which corresponds to the opening of a gap when entering in the ordered state \cite{Onuki1992, Troc2003, Troc2004, Troc2006}. Similar anomalies in $\rho (T)$ appear e.g.~in Cr at its spin density wave transition \cite{Fawcett1988}, or in URu$_2$Si$_2$ at the hidden order transition \cite{Hassinger2008}. For a current applied along the $a$ direction a sharp kink at $T_C$ indicates the onset of the ferromagnetic order due to the suppression of the spin disorder scattering (see below Fig.~\ref{fig2}(d)).  At $T^\star$, no clear anomaly is detected, however, the derivative $d\rho / dT$ shows a broad maximum at $T^\star$ for both current directions as shown in the inset of Fig.~\ref{SvsSdHvsCvsT} (b). This criterion had been used previously to determine $T^\star$ as function of pressure \cite{Oomi1995}.

Both, the thermoelectric power and the resistivity, are much higher for heat or charge current along the $b$ axis compared to the $a$ axis. This is related to the strong anisotropy of the Fermi surface with cylindrical Fermi surfaces along the $b$ axis \cite{Satoh1992, Settai2002, Samsel-Czekala2011}. The sign change in the thermoelectric power for heat current along the $a$-axis suggests that the main heat carrier are changing at the cross-over at $T^\star$ from electron above $T^\star
$ to hole-like heat carriers below $T^\star$. A similar change of the charge carriers has been reported from Hall effect experiments \cite{Tran2004}. Assuming a simplified one band model for the Hall effect analysis, at $T=2$~K the hole concentration reaches nearly 2 holes/f.u. With increasing temperature this concentration decreases rapidly. At high temperatures a carrier number of 0.4 electrons/f.u. has been found ($T > 160$~K).\cite{Tran2004}

In specific heat divided by temperature ($C/T$) (Fig.~\ref{SvsSdHvsCvsT}(c)) a mean-field like the second order phase transition appears at $T_C$. At $T^\star$ a broad anomaly indicates the cross-over from the weakly polarized FM1 to the strongly polarized FM2 state.  This anomaly gets more pronounced after subtracting the phonon contribution to the specific heat as has been demonstrated in detail in Refs.~\onlinecite{Hardy2009, Troc2012}.

Comparing the signatures of the cross-over in the different experimental probes,  TEP gives a very pronounced anomaly at $T^\star$ for both heat current directions, $\nabla T \parallel a$ and $\nabla T \parallel b$. It is interesting to note that the thermal expansion is very anisotropic in UGe$_2$ with an increase of the linear thermal expansion coefficient $\alpha _a$ along the $a$ axis while sharp negative anomalies appear for $\alpha _b$ and $\alpha _c$.\cite{Hardy2009} In contrary to the specific heat the thermal expansion exhibit large anomalies at the cross-over $T^\star$ while the characteristic cross-over temperature depends on the crystal direction. A well-defined signature of a phase transition will occur for a pressure above the CEP.

Under magnetic field the cross-over temperature $T^\star$ is increasing with magnetic field applied along the easy magnetization $a$ axis	as shown in Fig.~\ref{fig2}(a) for $H =9$~T. Fig.~\ref{fig2}(b) shows a zoom to the low temperature regime  of the thermoelectric power. Following the Boltzmann picture in which the TEP is given by the Mott formula
\[ 
S=-\dfrac{\pi^2}{3} \dfrac{{k_B}^2T}{e}
 \left(  \dfrac{\ln\sigma(\epsilon)}{\partial \epsilon} \right)  _{\epsilon_F}
\]
where $k_B$ is the Boltzmann constant, $-e<0$ is the electronic charge. $\sigma(\epsilon)$ is the electric conductivity of the system as a function of energy and $\epsilon_F$ is the Fermi energy \cite{Mott1958, Ziman1960, Behnia2004}. From this it is obvious that TEP depends on both, the transport and the thermodynamic properties of the system. In addition, the total TEP in a multiband system is the sum of the TEP of each band weighted by its respective electrical conductivity \cite{Miyake2005}. This makes a detailed analysis of the temperature dependence of the TEP often complicated. 

In a single band approach in the low temperature limit, the TEP is found to be correlated to the electronic specific heat coefficient $\gamma$  in the Fermi liquid regime and a dimensionless $q$-factor $q=S N_A e/(\gamma T) \approx \pm 1$ has been observed in a wide range of strongly correlated electron systems \cite{Behnia2004}. Here $N_A$ is the Avogadro number and $e>0$ is the elementary electronic charge with the entropy defined as $S_e=\gamma T$  \cite{Behnia2004}. This shows that strong renormalization effects in the TEP and the Sommerfeld coefficient $\gamma$ cancel each other out in spite of the different physical origin of these thermodynamic quantities. This ratio characterizes the thermoelectric materials in terms of an effective charge carrier concentration per formula unit \cite{Behnia2004, Miyake2005, Zlatic2007}, thus it is inversely proportional to the number of heat carriers per formula unit. Fig.~\ref{fig2}(c) shows $S(T)/T$ below 2.5~K at zero field. We estimated the value of $S(T)/T$ in the limit $T\rightarrow0$ to $S/T=0.24\mu$VK$^{-2}$; the specific heat coefficient $C/T |_{T \to 0}$ extrapolates to $\gamma = 33.2$~mJmol$^{-1}K^{-2}$.\cite{Lashley2006} With these values we evaluate a $q$-factor of $0.7$. This positive value is quite close to $1$ in the simplified approximation of a spherical Fermi suface despite the fact the UGe$_2$ is a multiband system with a complex Fermi surface as discussed below.

A strong field dependence of $S(T)$ can be observed, even below 5~K, in contrast to the specific heat, which is almost constant as a function of magnetic field \cite{Hardy2009}. This indicates that already in this temperature regime the scattering term is important to  evaluate the thermoelectric response. 

In Fig.~\ref{fig2}(d) we show the temperature dependence of the electrical resistivity. We compare $\rho (T)$ for electrical current $j \parallel b$ and $j\parallel a$ at $H=0$ and at 8~T applied along the easy magnetization axis $a$. As discussed above, in zero magnetic field, the signature at the ferromagnetic transition is different, with a hump for the transverse $j \parallel b$ and a sharp kink for the longitudinal configuration $j \parallel a$. The hump-like feature for $j\parallel b$ is also observed at the ferromagnetic cross-over under the applied field along $a$ while in the transverse configuration the resistivity decreases at the cross-over. The smearing of the anomalies for field along the easy magnetization axis is in excellent agreement with the drastid drop of the specific heat anomaly under field.\cite{Hardy2009}  No clear feature is associated to the cross-over at $T^\star$. 

In the transverse configuration under field $\rho (T)$ starts to increase already as the temperature decreases below 13~K due to strong orbital contribution to the resistivity, while the intrinsic longitudinal resistivity always decreases with decreasing temperature (Fig.~\ref{fig2}(d)). This increase of the transverse magnetoresistivity at low temperatures under applied field is a clear signature that UGe$_2$ is a compensated metal and shows that on cooling the condition $\omega _c \tau \gg 1$ will be fulfilled ($\omega _c$ being the cyclotron frequency and $\tau$ scattering rate), i.e.~the orbital contribution to the magnetoresistance will dominate.  

\subsection{Quantum Oscillations}

\begin{figure}[t!]
	\includegraphics[width=0.9\linewidth]{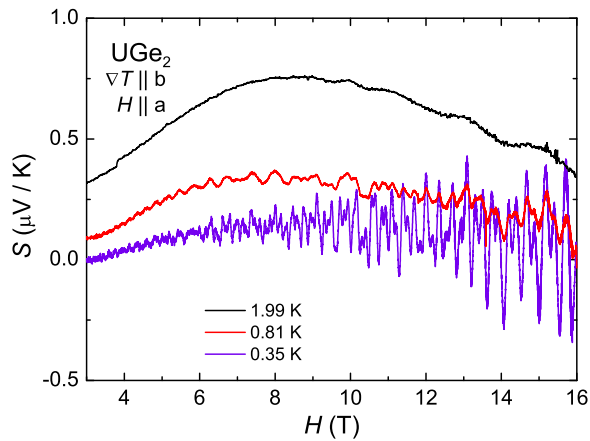}
	\caption{\label{SvsH} (Color online). Field dependence $S(H)$ for different temperatures above 3~T showing quantum oscillations (sample \#2).}
\end{figure}

The thermoelectric power as a function of magnetic field $S(H)$ for sample \#2 is plotted in Fig.~\ref{SvsH} for different temperatures in the field range from 3 to 16~T. The absolute value of $S$   at low temperature ($T < 1$~K) is very small, $S < 0.1 \mu$V/K. To obtain a significant thermal gradient of about 3\% between both ends of the sample requires increasing from $T_{min} \approx 100$~mK to almost 200~mK due to the large thermal conductivity. This excludes proper measurements below 200~mK. The quantum oscillations have been observed by sweeping the magnetic field along the $a$-axis with a constant rate of $0.1$~T/min.  In agreement with the large transverse magnetoresistance of the sample (see below) the thermal conductivity decreases strongly with field and thus the thermal gradient during the field sweep is not constant. The changing gradient has been taken into account for the $S (H)$ curves shown in Fig.~\ref{SvsH} where the field dependence of the isothermal thermopower for different temperatures between $T = 0.35$~K and $2$~K is plotted. At $T=2$~K $S (H)$ increases with field up to $H \approx 8$~T and decreases for higher fields. Even at this temperature quantum oscillations can be observed. On lowering the temperature the overall shape of the average TEP with a broad maximum at $H\approx 8$~T does not change, but the amplitude of the non-oscillatory part $S_{av}$ of the signal decreases. In contrast, the size of the oscillations of the TEP increases strongly at low temperatures. At the lowest temperature presented ($T \approx 350$~mK), quantum oscillations can be observed above $H \approx 3$~T. While the size of the average signal $S_{av}$ is almost zero, the oscillating part of the thermopower at the highest field measured, $H = 16$~T, reaches values of $S_{osc} \sim \pm 0.4 \mu$~V/K, i.e.~the value of the thermopower at 2~K. In contrast to other transport properties, $S_{av}$ strongly decreases for $T \to 0$~K. As $S$ measures the entropy per carriers $S$ should vanish for $T \to 0$ in a metal.

For sample \#1 qualitatively similar results have been observed. The amplitudes of the oscillations for this crystal are smaller, indicating the lower sample quality. Especially the thermal conductivity in sample \#1 is smaller allowing TEP experiment to somewhat lower temperatures than those in sample \#2. 

\begin{figure}[t!]
	\includegraphics[width=0.9\linewidth]{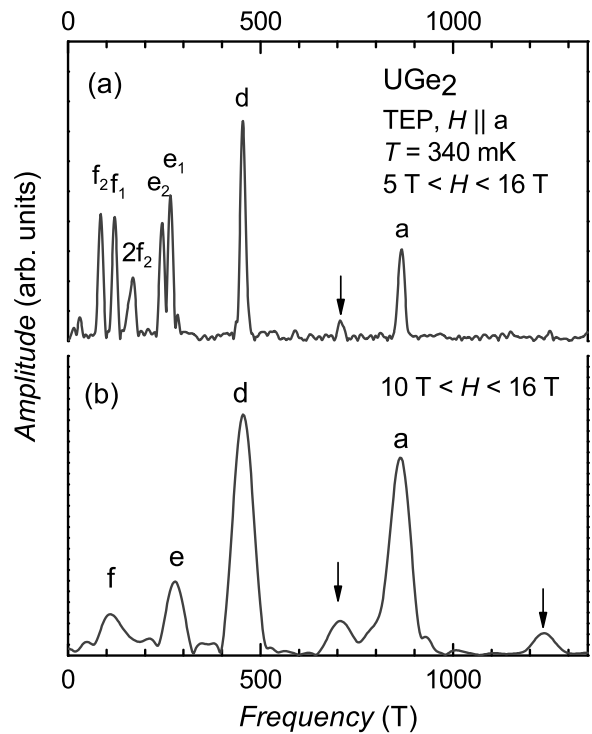}
	\caption{\label{TEP_FFT} (Color online). Fast Fourier transformation of $S(H)$ at $T = 350$~mK in the field range from (a) 5~T to 16~T and (b) 10.5~T to 16~T, respectively. The nomination of the Fermi surface branches follows that of Ref.~\onlinecite{Haga2002}. Arrows indicate branches which have not been observed previously.}
\end{figure}

In Fig.~\ref{TEP_FFT} (a) and (b) we show the Fourier spectrum of the TEP signal in the large field range from  5~T to 16~T and from  10.5~T to 16~T at 340~mK for sample \#2. For the FFT analysis we neglect the effect of the magnetization $M$ to the effective magnetic field magnetic field $B = \mu_0(H + (1-D)M)$, as the demagnetization factor $D \approx 1$.
A polynomial background to model the non oscillatory part $S_{av}$ has been subtracted from the measured $S (H)$ data. The obtained FFT spectra do not depend on the rank of the polynomial. The observed frequencies for sample \#2 are listed in Tab.~\ref{table} and we find a good agreement with the previous dHvA experiment for $H \parallel a$. In the spectra over the whole field range (Fig.~\ref{TEP_FFT} (a)) we detect at least five different Fermi surface orbits. The labeling of the frequencies follows that of Ref.~\onlinecite{Haga2002} with the orbits $a \approx 865$~T, $d \approx 454$~T; the low frequencies $e_1$, $e_2$ and $f_1$, $f_2$ can only be resolved in the low field range. In addition we observed in the TEP a branch at 710~T in sample \#2 and in the field range from 10 to 16~T another frequency at 1235~T which had not been observed in the previous dHvA experiments \cite{Terashima2002, Haga2002}.  Contrary, the branch called $c \approx 980$~T in the dHvA experiments  has not been detected in our TEP and SdH experiments, neither in samples \#1 nor \#2.  
 
\begin{figure}[t!]
	\includegraphics[width=0.9\linewidth]{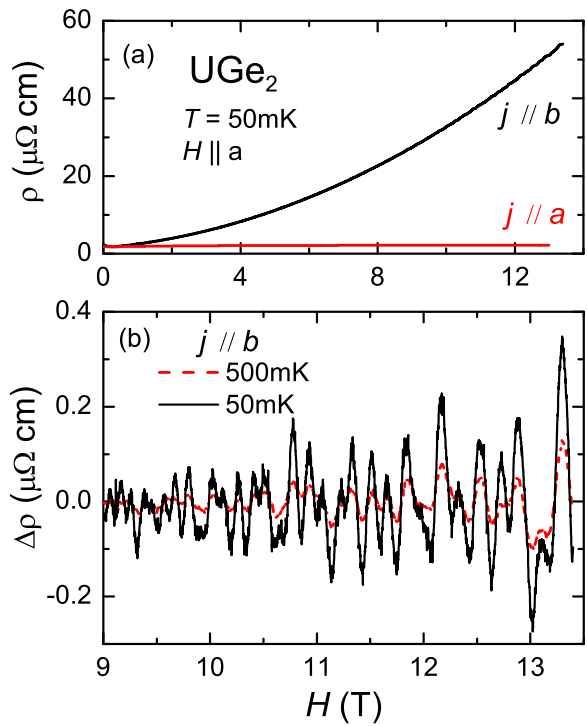}
	\caption{\label{SdH} (Color online). (a) Transverse and longitudinal magnetoresistance of UGe$_2$ (sample \#2) for $H\parallel a$ and $j \parallel b$ axis at $T = 50$~mK and $j \parallel a$. (b) Zoom on the oscillatory part of the magnetoresistance for $j \parallel b$ at 50~mK and 500~mK in the field range from 9 to 13.4~T, determined from the raw data after subtracting a polynomial background.}
\end{figure}

\begin{figure}[h]
	\includegraphics[width=0.9\linewidth]{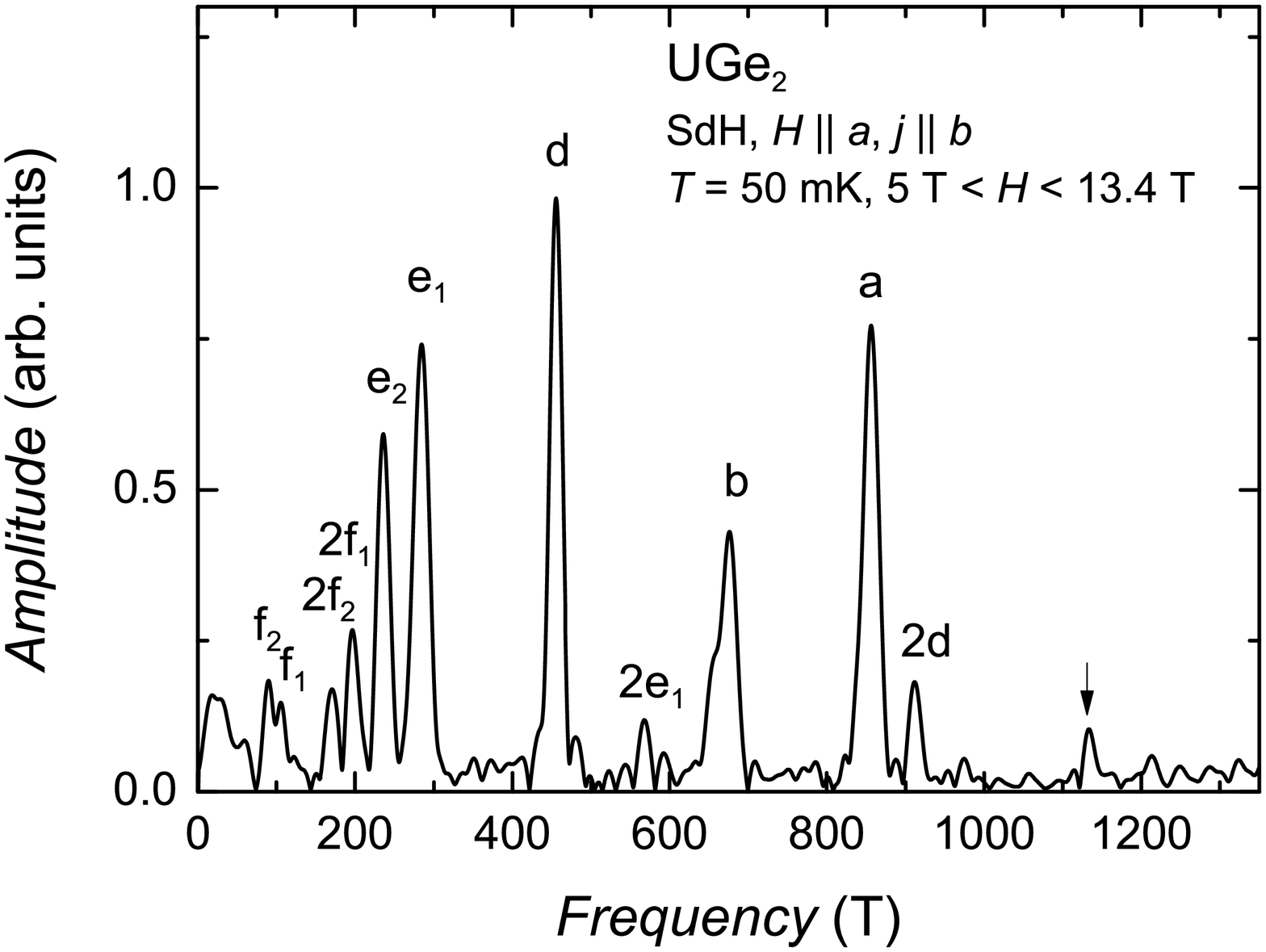}
	\caption{\label{SdH_FFT} (Color online) Fast Fourier transformation spectrum of the Shubnikov de Haas oscillations at 50~mK for the field range from 5.5~T to 13.4~T for field along the $a$-axis. }
\end{figure}

\begin{figure}[h]
	\includegraphics[width=0.9\linewidth]{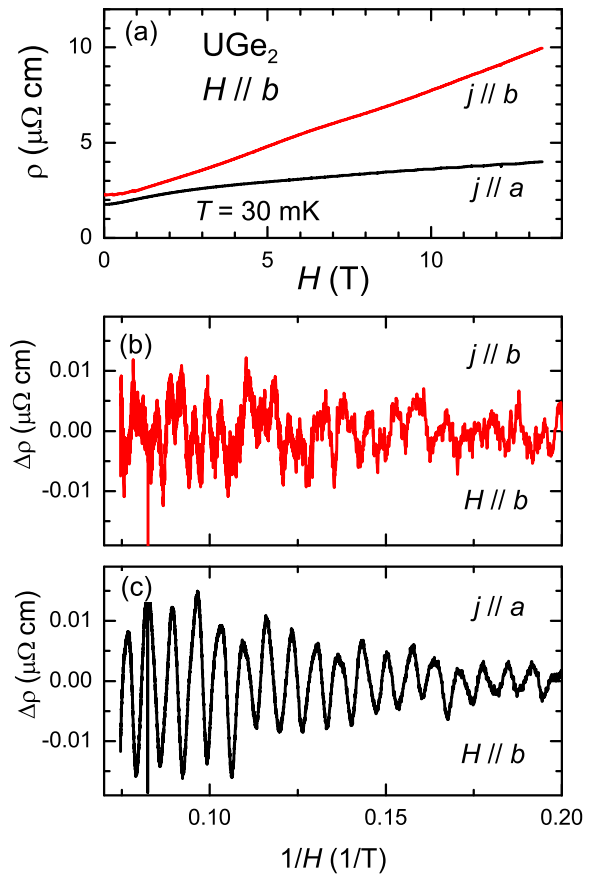}
	\caption{\label{SdH_parallel_b} (Color online) (a) Transverse and longitudinal magnetoresistance for field along the $b$-axis.  The oscillatory part of the resistivity is shown in (b) for the longitudinal and (c) transverse configuration of the resistivity for field applied along the $b$ axis. }
\end{figure}

\begin{figure}[h]
	\includegraphics[width=0.9\linewidth]{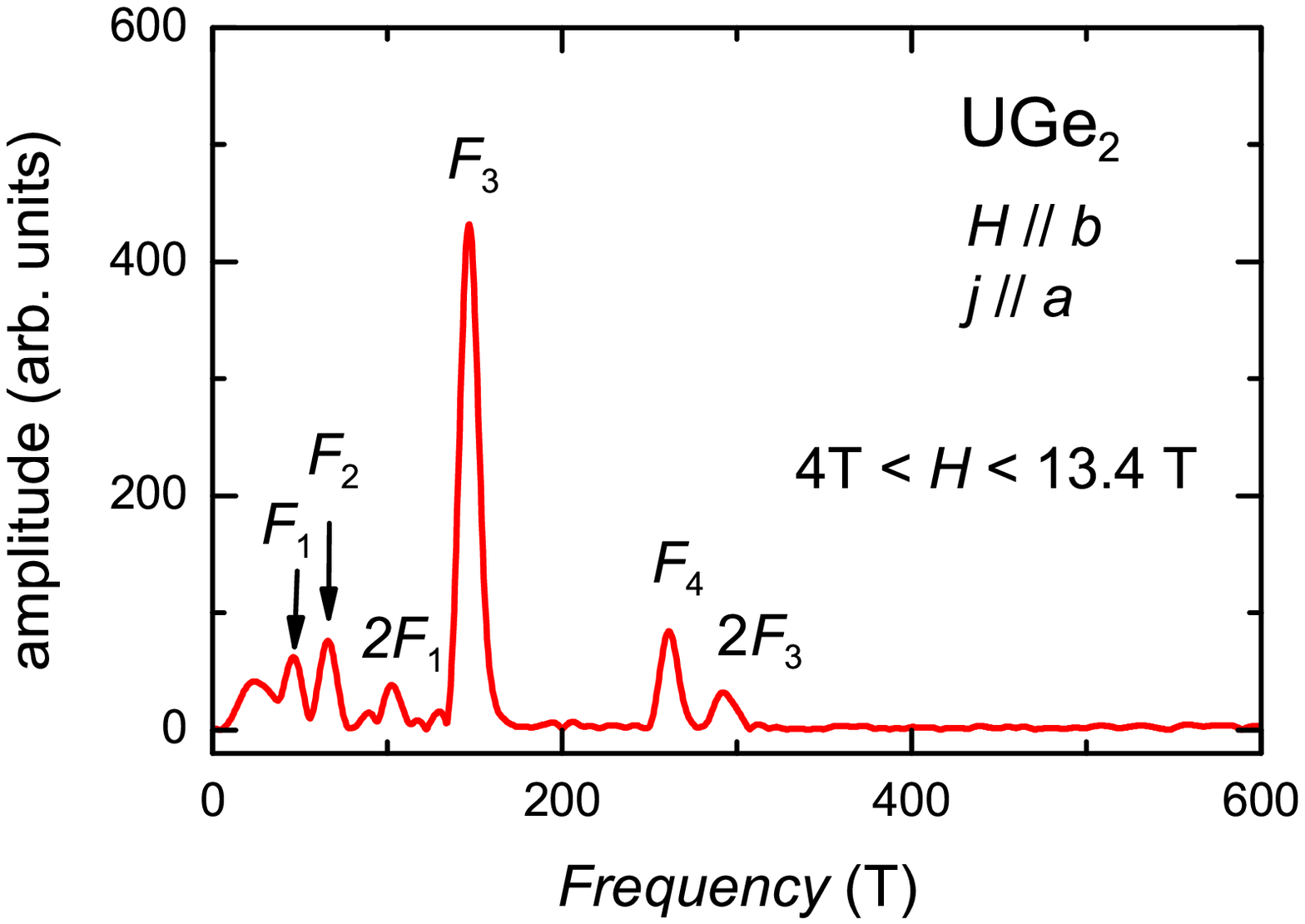}
	\caption{\label{FFT_b-axis} (Color online) Fast Fourier transformation spectrum of the Shubnikov de Haas oscillations at 50~mK for the field range from 4~T to 13.4~T for field along the $b$-axis. }
\end{figure}

In Fig.~\ref{SdH}(a) we show the transverse ($j \parallel b$) and longitudinal  ($j \parallel a$) magnetoresistance of sample \#2 at $T = 50$~mK as function of field applied along the $a$-axis. The transverse magnetoresistance is extremely large and positive $\Delta \rho_\perp / \rho_0 = \rho ({\rm 13 T} - \rho (H=0)/ \rho (H=0) =  24.5$ except at lowest field ($H < 0.25$~T). 
Above 4.5~T the magnetoresistance has a $H^2$ field dependence indicating that UGe$_2$ is a compensated metal and that in the high field region the magnetoresistance is determined by the orbital motion of the quasiparticles at the Fermi surface, hence $\omega_c \tau \gg 1$. \cite{Pippard1989} Contrary, the longitudinal magnetoresistance is small, $\Delta \rho_\parallel / \rho (H=0) = 0.24$ for a field of 13~T.  It starts to  saturate above 2~T in agreement with the absence of the orbital motion of the electrons. 

\begin{table*}[<centered>]
\begin{ruledtabular}
\caption{\label{table} Comparison of quantum oscillation frequencies and the cyclotron masses for $H\parallel a$ axis of  of UGe$_{2}$ (sample \#2) from TEP following eq.~\ref{Young}, SdH and previous dHvA measurements taken from Refs.~\onlinecite{Haga2002, Terashima2002}.}
	\begin{tabular}{ccccccccc}

		\multicolumn {1}{c}{ } &\multicolumn {2}{c}{TEP} & \multicolumn{2}{c}{SdH} & \multicolumn{2}{c}{dHvA\cite{Haga2002} } & \multicolumn{2}{c}{dHvA\cite{Terashima2002} }\\
		\multicolumn {1}{c}{ } &\multicolumn {2}{c}{(5--16~T)} & \multicolumn{2}{c}{(5--13.4~T)} & \multicolumn{2}{c}{(5--17~T)} & \multicolumn{2}{c}{(5--18~T)} \\
  		\multicolumn {1}{c}{ } & $F$~(T) & $m^\star$~($m_0$) & $F$~(T) & $m^\star$~($m_0$) &$F$~(T) & $m^\star$~($m_0$)  &$F$~(T)& $m^\star$($m_0$) \\
		\hline
  		 f2 & 85 & 1.6 & 89 & - &   &  & 92.1 & \\  		
  		f1 & 120 & 2.31 & 105 & 1.08 & 97 & - & 106.1 &\\
  		e2 & 245 & 4.16 & 236 & 3.6 & 257 & 4.2 & 259.1 & 5.0(8)\\
  		e1  & 263 & 4.6 & 284 & 3.9 &  &  & & \\
  		d & 455 & 4.6 & 453 & 4 & 434 & 5.4 & 439.5 & 4.0(7) \\
  		b &   &   & 671 & 4.6 & 661 & 4.0 & 671.6 & \\
  		 & 710 & 7.0 &   &   &   &  & & \\
  		a & 863 & 6.2 & 855 & 5.3 & 852 & 4.8 & 862 & 5.4(4)\\
  		c & 1010 &   &   &   & 980 & 5.3 & 992.1 & \\
  		  &   &   & 1129 & - &   &  & &\\
  		  & 1237 & 4.0 &  &   &   &  & &\\
	\end{tabular}
	 
\end{ruledtabular}	   
\end{table*}

Shubnikov de Haas oscillations for $H \parallel a$ can be resolved for magnetic fields above 5.5~T in the transverse configuration. However, even at the highest field they are small compared to the non-oscillatory part of the magnetoresistance. This is in strong contrast to the TEP oscillations discussed above. The FFT spectrum for the field range from 5.5~T to the highest field of 13.4~T of the magnetoresistance at 50~mK is shown in Fig.~\ref{SdH_FFT} and the obtained frequencies are also given in Tab.~\ref{table}. Due to an imperfect alignment of the sample in both experiments, TEP and SdH respectively, the obtained frequencies do not match perfectly. However, all previously observed main frequencies have been detected in the spectrum of SdH. We clearly observe the splitting of the small Fermi surface orbits f and e, as well as the harmonics of these. Another orbit at $F = 1133$~T has been detected, however, for higher temperatures it cannot be resolved anymore. 
Comparing thermopower quantum oscillations and SdH we notice: (i) In both probes we can detect the same frequencies as previously seen in the dHvA experiment. (ii)
The amplitude of the oscillations compared to the non-oscillatory signal is by far higher in the TEP. (iii) In the TEP quantum oscillations are observed up to at least 2~K while in SdH the oscillations vanish at about 600~mK as can be seen in Figs.~\ref{SvsH} and \ref{SdH}.

In Fig.~\ref{SdH_parallel_b}(a) we show the magnetoresistance for magnetic field $H \parallel b$ for the longitudinal $j \parallel b$ and transverse  $j \parallel a$ configuration. Astonishingly, the longitudinal magnetoresistance in this configuration is larger than the transverse ($j \parallel a$) which tends to saturate above 4~T. In general the transverse magnetoresistance is strongly influenced by the Fermi surface topology, and the observed behavior is in agreement with the open orbits formed along the $b$ axis \cite{Onuki1991}, no saturation of the transverse magnetoresistance has been reported at higher temperatures in Refs.~\onlinecite{Oomi1995, Troc2003, Troc2004, Troc2006}. Fig.~\ref{SdH_parallel_b}(b-c) shows the oscillatory part of the magnetoresistance as function of inverse magnetic field $1/H$. Figure~\ref{FFT_b-axis} shows the FFT analysis of oscillations in the transverse configuration in the field range from 4 to 13.4~T. We observe at least four different frequencies, $F_1 \approx 48$~T, $F_2 = 66$~T, $F_3 = 146$~T, and $F_4 = 261$~T with effective masses of $1.6 m_0$, $1.2 m_0$, $2.4 m_0$, and $3.3 m_0$, respectively. These small orbits have been reported in dHvA  \cite{Satoh1992}. However, the large cyclotron orbits  could  not be observed in the SdH oscillations. 

Quantum oscillations are a powerful tool for probing the Fermi surface properties and to obtain microscopic information like the effective mass of the electrons or their mean free path. As mentioned above the standard experimental probes to observe magnetic quantum oscillations is the measurement of the magnetization which is directly related to the free energy and probes the oscillations of the density of states at the Fermi surface. The oscillations in this case are excellently described by the well known Lifshitz-Kosevich theory which is based on a thermodynamic approach \cite{Shoenberg1984}. On the contrary, resistivity and thermoelectric power are transport properties and the electronic scattering has to be taken into account. Furthermore, the TEP is measured by the application of a thermal gradient. Quantum oscillations in the SdH effect\cite{Adams1959} can be understood taking Pippards argument \cite{Pippard1965} into consideration that the scattering probability is proportional to the number of states in which electrons can be scattered and thus to the density of states at the Fermi level $D(\epsilon_F)$. Therefore
\[\frac{\tilde{\sigma}}{\sigma} = \frac{\tilde{D}}{D}\]
 which connects the oscillatory part of the conductivity $\tilde{\sigma}$ with the oscillatory part of the density of states $\tilde{D}$. Thus, within a reasonable approximation the amplitude $A(H,T)$ of the oscillations in the SdH is given by the Lifshitz-Kosevich  formula \cite{Shoenberg1984}.

\begin{equation}
\label{LK_all}
A (H,T) \propto H^{1/2}\left|\frac{\partial^2 S}{\partial k^2}\right|^{-1/2}R_T R_D R_S
\end{equation}

\noindent with the temperature damping factor

\begin{equation}
\label{LK}
R_T = \frac{\alpha p X}{\sinh(\alpha p X)}\;, 
\end{equation}

\noindent where $\alpha = 2 \pi^2 k_B/e \hbar$, $p$ is the number of the harmonics, and $X=m_c^\star T/H$ which allows for the determination of the effective mass $m_c^\star$.  
From the Dingle damping factor
\[
R_D = \exp\left(-\frac{\alpha p m_c^\star T_D}{H}\right)
\]
the Dingle temperature $T_D = \hbar/2\pi k_B \tau$ can be determined and thus microscopically the mean free path. $R_S = \cos (\frac{\pi g p m_c^\star}{2m_0})$ is the spin splitting term due to the Zeeman splitting of the Landau level. 

In contrast, quantum oscillations in the thermoelectric power are treated differently in the literature.
In Ref.~\onlinecite{Kirichenko2008} the thermoelectric coefficients have been calculated by Kirichenko \textit{et al.}~in the presence of a thermal gradient and  for the temperature dependence of the oscillation amplitude the following expression for the temperature dependent term of the amplitude of the quantum oscillations has been given:
\begin{equation}
\label{Peschanskii}
A(T) = -\frac{3}{\alpha X}\frac{\sinh(\alpha X)-(\alpha X)\cosh(\alpha X)}{\sinh^2(\alpha X)}
\end{equation}
In this model the amplitude goes to 1 as $T \to 0$. As we will see below, it has exactly the same $T$ dependence of the amplitude as Lifshitz-Kosevich. 

As pointed out in Refs.~\onlinecite{Young1973, Fletcher1981, Fletcher1983} by Young and later by Fletcher the TEP depends on the derivative of the density of states at the Fermi level. Independently, in Ref.~\onlinecite{Pantsulaya1989} by Pantsulaya and Varlamov  the amplitude of the oscillations of the longitudinal thermoelectric coefficient has been calculated taking into account the energy dependence of the electrons relaxation. As the oscillations of the conductivity are in general small compared to those in the thermoelectric coefficient, the behavior of the thermopower coincides with that of the thermoelctric coefficient \citep{Pantsulaya1989}. The temperature dependence of the thermoelectric power oscillations obtained in these Refs.~\onlinecite{Young1973, Fletcher1981, Pantsulaya1989}  is given by the derivative of the Lifshitz-Kosevich formula, 

\begin{equation}
\label{Young}
A (T) \propto \frac{(\alpha p X)\coth(\alpha p X)-1}{\sinh(\alpha p X)}\;.
\end{equation}

Thus, the amplitude of the TEP quantum oscillation will show a maximum for $\alpha p m_c^\star T/H = 1.62$,\cite{Young1973} i.e. at 
\begin{equation}
T\approx 0.11 H/(p m_c^\star) \quad ,
\end{equation}
and will vanish for $T \to 0$~K, in contrast to the LK formula, where the amplitude would be maximal at $T = 0$~K, which is non-physical, as the entropy has to vanish for $T \to 0$~K. As the TEP, at least in the low temperature limit, gives the entropy per charge carrier, one would expect that at $T=0$ the TEP vanishes, which is true for the non-oscillating average signal $S_{av}$.

To determine the cyclotron masses $m_c^\star$ of each branch, we have analyzed the temperature dependence of the amplitudes of the FFT spectra for each branch following the different approaches in the field ranges from 5~T to 10~T and from 10~T to 16~T. No strong field dependence of the masses had been observed, but as can be seen in Fig.~\ref{TEP_FFT} the low frequencies are only resolved in the low field range while the large frequencies appear only in the high field range. The effective field over the field window of the FFT is given by $\frac{1}{H_{eff}}=\frac{1}{2}(\frac{1}{H_{min}}+\frac{1}{H_{max}})$. 
 As explained above, the temperature and  the temperature gradient was not constant during the field sweep. We corrected this by averaging the temperature over the corresponding field window of the FFT analysis. 

\begin{figure}[t]
	\includegraphics[width=1\linewidth]{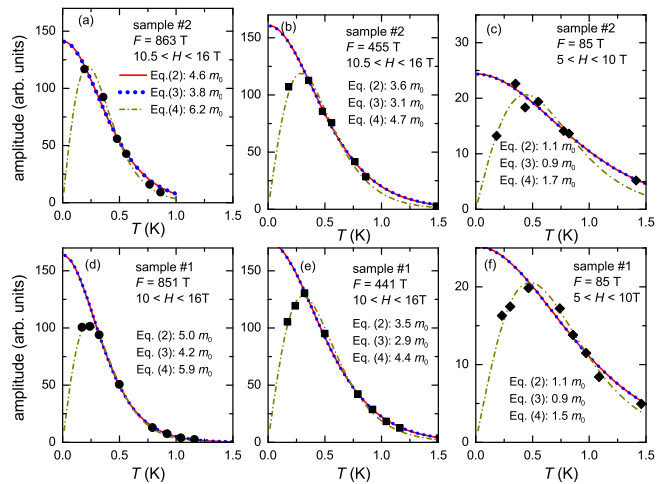}
	\caption{\label{FFT_amplitude} (Color online) Temperature dependence of the FFT amplitudes for different branches for sample \#2 (a-c) and sample \#1 (d-f). The lines are least square fits following the Lifshitz-Kosevich formula, Eq.~(\ref{LK}) (red solid lines), the derivative of the Lifshitz-Kosevich formula following Young, Fletcher, and Pansulaya and Varlamov, Eq.~(\ref{Young}) (Refs.\onlinecite{Young1973,Fletcher1981,Pantsulaya1989}, dark dashed dotted lines), and the formula of Kirichenko \textit{et al}.~Eq.~(\ref{Peschanskii}) (Ref.~\onlinecite{Kirichenko2008}, blue dotted line). For fits following Eqs.~\ref{LK} and \ref{Peschanskii} the lowest temperatures have not been taken into account. The respectively obtained effective masses are also indicated. }
\end{figure}
 
Figure \ref{FFT_amplitude} (a-c) shows the temperature dependence of the FFT amplitudes of the main frequencies for sample \#2 and (d-f) for sample \#1. The behavior  for both samples is very similar. The differences in the frequencies is due to slightly different orientations in the field. The amplitude has been fitted with the different models introduced above. It is obvious that the amplitude is decreasing  at the lowest temperature for all shown frequencies. Due to this the fit following the Lifshitz-Kosevich formula Eq.~(\ref{LK}) or Kirichenko \textit{et al}.\cite{Kirichenko2008}, Eq.~(\ref{Peschanskii}) cannot reproduce the experimental data to the lowest temperature. Interestingly both, Eqs.~(\ref{LK}) and (\ref{Peschanskii}) result in exactly the same temperature dependence while the extracted mass following Eq.~(\ref{Peschanskii}) is about 80\% of that from the Lifshitz-Kosevich formula. The temperature dependence of the experimentally observed amplitudes are in good agreement with Eq.~(\ref{Young}) which is the derivative of the Lifshitz-Kosevich term. The high temperature tail of the amplitude for all frequencies is very well fitted by the three models. The extracted masses from the different models are also indicated in Fig.~\ref{FFT_amplitude}. The mass obtained by Eq.~(\ref{Young}) is by a factor of about 1.4 and 1.6 higher than that obtained from Eqs.~(\ref{LK}) and (\ref{Peschanskii}), respectively. However, the crucial factor for the determination of the mass are data at lowest temperature. Only for the lowest temperatures the deviation from the Lifshitz-Kosevich formula and the decrease of the oscillation amplitudes have been observed. In Tab.~\ref{table} we have listed the effective masses obtained from the analysis of the TEP, SdH, and from the previous dHvA experiments \cite{Haga2002, Terashima2002}.

In Refs.~\onlinecite{Fauque2013, Lin2013} the authors have used the Lifshitz-Kosevich formalism to analyze the quantum oscillations in the Nernst coefficient divided by temperature $\nu / T$ in Bi$_2$Se$_3$ and doped SrTiO$_3$. As discussed above, following the Mott formula the thermoelectric power is the sum of two term, one depending on the density of states while a second term including the electronic scattering. As in the low temperature limit the thermoelectric power $S/T$ is eventually correlated to the specific heat coefficient $\gamma$, it seems plausible to use the Lifshitz-Kosevich formalism. Indeed in Refs.~\onlinecite{Fauque2013, Lin2013} a very good agreement between the effective masses determined from the analysis of the oscillations in $\nu / T$ and SdH quantum oscillations have been observed. We did an analogue analysis of the observed thermoelectric power quantum oscillations in UGe$_2$ and found reasonable agreement between that analysis and the masses determined from SdH, or the previously dHvA oscillations. However, a sound theoretical justification of such an approach is still missing.

\subsection{Band-structure Calculations}

Previous band-structure calculations \cite{Shick2001, Settai2002, Yaresko2004, Samsel-Czekala2011} for UGe$_2$ indicate that the main frequencies of the Fermi surface are the large cyclotron orbits of the cylindrical Fermi surface along the $b$ axis. For $H \parallel b$ three different dHvA frequencies at $F_\gamma =6860$~T, $F_\alpha = 7760$~T, and  $F_\beta = 9060$~T with large effective masses (23~$m_0$, 15~$m_0$, and 18~$m_0$, respectively) have been reported in the dHvA experiment. As discussed above, these large orbits could not be observed in the magnetoresistance, but only several  small frequencies below 300~T. 

\begin{figure}[h]
	\includegraphics[width=0.95\linewidth]{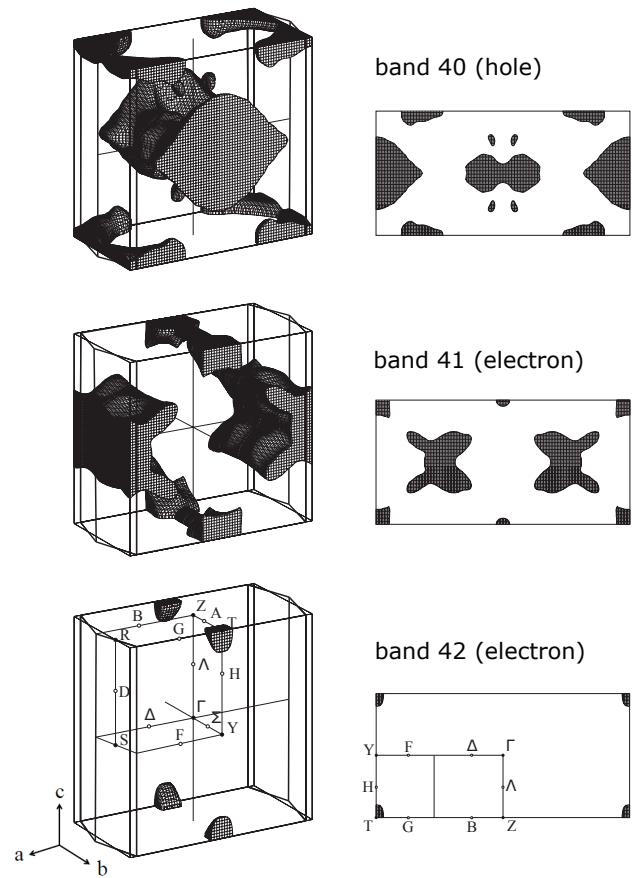}
	\caption{\label{Fermi_surface} Calculated Fermi surface of UGe$_2$ in the ferromagnetic state. The Fermi surfaces are centered at the $\Gamma$-point of the Brillouin zone (left row) and drawn in the primitive orthorhombic Brillouin zone. The bold lines show the Brillouin zone in the correct $c$-based-centered orthorhombic $k$-space. The volume of both Brillouin zones is equal. The high symmetry point are indicated for band 42.  (Right row) Cross sectional surfaces of the Fermi surfaces in the $ac$ plane according to the symmetry points indicated in the Fermi surface of band 42. }
\end{figure}

Fig.~\ref{Fermi_surface} shows the newly calculated Fermi surface of UGe$_2$ using the Dirac-based LAPW method \cite{Yamagami2000} with a local-spin-density approximation  (LSDA) including orbital polarization interaction  $I \beta \langle \ell \rangle \hat{\ell}$ along the $a$ axis plotted in the primitive orthorhombic Brillouin zone, where $\beta$ is the well-known relativistic operator, $\langle\ell\rangle$  and $\hat{\ell}$ is an average value and an operator of orbital angular momentum, respectively, and the orbital parameter $I$ is set to 0.002 Ry/$\mu_{\rm B}$. We have also plotted the correct $c$-based-centered orthorhombic zone which has same volume. The calculated Fermi surfaces have somewhat different shapes than those of the previous calculation \cite{Settai2002}. This is mainly due to the accuracy of the numerical calculation on mesh points of Brillouin zone to make up the Fermi surfaces, acually using total 1512 points with 14$\times$8$\times$14 mesh of the primitive orthorhombic Brillouin zone. The obtained magnetic moment calculated from all the occupied states  of $1.3 \mu_B$ is very close to the experimentally observed moment of $1.4 \mu_B$ in the FM2 state.  The calculated dHvA frequencies and effective masses are given in Tab.~\ref{table2}. We also plotted in Fig.~\ref{Fermi_surface} the cross-sectional surfaces of Fermi surfaces in the $ac$ planes. Here it should be noted that the  area of Fig.~\ref{Fermi_surface} is twice as that of the $c$-based-centered orthorhombic zone.
 
In contrast to the previous calculation \cite{Settai2002} we find that only three bands cross the Fermi surface, the  band 37 of the previous calculation goes to the fully occupied state. Band 40 is a hole Fermi surface and the bands 41 and 42 are electron Fermi surfaces. The obtained frequencies for the main cylindrical frequencies along the $b$ axis are somewhat smaller than in the previous calculation and also smaller than the experimentally observed dHvA frequencies. Also  the high effective masses are not completely reproduced. The calculated Sommerfeld coefficient of the specific heat is 15.6 mJ~mol$^{-1}$K$^{-2}$ by comparison to the experimental value of 60~mJ~mol$^{-1}$K$^{-2}$. Naturally, as it happens for many heavy fermion materials, the calculated cyclotron masses are smaller than the observed ones, considering the enhancement factor between the experimental and calculated specific heat coefficients, the magnitude of the cyclotron masses are in the same order. While a comparison with the calculated Fermi surface along the $b$ axis with the main cylindrical Fermi surfaces shows certain similarities, for the $a$ axis this is not the case. 

\begin{table}[<centered>]
\begin{ruledtabular}
\caption{\label{table2} Calculated dHvA frequencies and cyclotron masses along the $b$ and $a$ axis for UGe$_2$}
	\begin{tabular}{ccccc} 
  
		\multicolumn {1}{c}{ } &\multicolumn {2}{c}{$H\parallel b$} & \multicolumn{2}{c}{$H\parallel a$} \\
		\multicolumn {1}{c}{ } & $F$~(T) & $m^\star$~($m_0$) & $F$~(T) & $m^\star$~($m_0$) \\
   		\hline
  		band 40 	& 3360	& 5.67	& 90	& 0.66 	\\
  					& 6610	& 3.28	&	 	& 		\\
  					& 1110	& 2.02	&		&		\\
  		band 41		& 4130	& 6.43	&		&		\\
  		band 42		& 510	& 0.65	& 400	& 0.97	\\
	\end{tabular}

\end{ruledtabular}	   
\end{table}

Our new result is in qualitative agreement with the previous band-structure calculations where also  three Fermi surface sheets have been reported \cite{Shick2001, Samsel-Czekala2011} with a quasi-two dimensional hole Fermi surface being open along the $b$ axis.
In the present calculation only two extreme orbits are found in the calculated Fermi surface for $H \parallel a$, which appear as small pockets in the band 40 and 42 Fermi surfaces at frequencies $F = 90$~T and 400~T. These frequencies are not far from the values of the experimentally observed branches $f_2 \sim 85$~T and $d \sim 440$~T. In the calculation presented in Ref. \onlinecite{Samsel-Czekala2011} several small closed Fermi surface pockets appear in this hole Fermi surface which may be identified with the frequencies observed in the present experiment (see Tab.~\ref{table}) for $H\parallel a$ and $H\parallel b$. However, an assignment and detailed comparison to the experimentally observed orbits is not possible as no predictions of the frequencies has been published.  

To calculate the  correct Fermi surface of this low symmetry ferromagnet UGe$_2$ appears clearly to be extremely difficult and details of the Fermi-surface topology could not be reproduced in the present calculation. Furthermore the band structure calculations with the LSDA including the phenomenologically-postulated orbital polarization cannot describe the origin of the weakly polarized FM1 and strongly polarized FM2 phases in UGe$_2$, though it were improved to the value of the magnetic moment. In a realistic way, we should develop a computable orbital-dependent potential based on the density-functional theory in order to study the correct topology of Fermi surface as well as the magnetic properties of UGe$_2$, and also  the other ferromagnetic uranium compounds. To achieve a complete theoretical description of all features of the $p$-$T$-$H$ phase diagram of  UGe$_2$, the determination of the Fermi surfaces in the different phases under pressure is essential.\citep{Wysokinski2015} Surprisingly, there is clear evidence of a Fermi surface change on entering in the FM1 phase above $p_x$, but up to now no calculation has been made to predict the new FS topology of the FM1 phase. 

\section{SUMMARY}
 In conclusion, we have been able to observe thermoelectric power and Shubnikov de Haas quantum oscillations  in the ferromagnetic superconductor UGe$_2$ for field along the easy magnetization $a$ axis. In comparison to SdH the oscillations in the thermoelectric power are extremely large and can be followed to temperatures almost four times higher. The observed orbits in both probes are in good agreement with previous dHvA experiments. The analysis of the temperature dependence of the  oscillation amplitudes in the thermoelectric power allows the determination of the effective mass of the charge carriers. The amplitude of the oscillations in the TEP shows a maximum as function of  temperature and vanishes for $T \to 0$, in agreement with the expectation that the entropy of the charge carriers should vanish at $T =0$. The position of the maximum depends on the value of the effective mass and the ratio of temperature divided by the effective magnetic field. 
 
 Thus thermoelectric power is a powerful tool to observe quantum oscillations in strongly correlated electron systems. Compared to SdH or dHvA experiments the maximal amplitude of the oscillations is not observed at lowest temperature which allows a determination of Fermi surface parameters even at rather elevated temperature in strongly correlated 
electron systems. 

New band structure calculations of UGe$_2$ in the ferromagnetic state  are presented; however the quantitative agreement with the experimentally observed quantum oscillations for $H \parallel a$ axis is still not fully satisfying and indicates the difficulties of the band structure calculation in the low symmetry orthorhombic crystal. Let us notice that in the other ferromagnetic superconductors URhGe and UCoGe only very few orbits have been detected \cite{Yelland2011, Aoki2014a} and thus a comparison with band structure calculations is actually outside careful analysis. UGe$_2$ remains the best material to follow the possible improvement of bandstructure calculations for these orthorhombic ferromagnetic examples.

\begin{acknowledgments}

We acknowledge K.~Behnia and B.~Fauqu\'e for insightful discussions and critical reading of the manuscript. H.~Harima, K.~Izawa,  V.~P.~Mineev, and A.~A.~Varlamov are acknowledged for fruitful discussions and T.~Terashima for communicating the exact values of the dHvA frequencies and masses of Ref.\onlinecite{Terashima2002}. 

We acknowledge the financial support from the French ANR (within the programs  PRINCESS, SINUS, CORMAT), the ERC starting grant (NewHeavyFermion), and the Universit\'{e} Grenoble-1 within the Pole SMINGUE.

\end{acknowledgments}


%

\end{document}